\documentclass{jfm}
\usepackage{epsfig,psfrag}
\usepackage{amsmath}
\usepackage{amsfonts}
\usepackage{mathrsfs}
\usepackage{amssymb}
\usepackage{bm,bigints}
\usepackage{color}
\usepackage{natbib}
\usepackage{mdwlist}
\usepackage{xr,url}

\begin{document}

\newcommand\beq{\begin{equation}}
\newcommand\eeq{\end{equation}}
\newcommand\beqa{\begin{eqnarray}}
\newcommand\eeqa{\end{eqnarray}}
\newcommand{\Sy}{{\cal S}}
\newcommand{\U}{{\cal U}}
\newcommand{\K}{{\cal K}}
\newcommand{\T}{\mathsfi {T}} 
\newcommand{\bx}{\mathbf{x}}
\newcommand{\by}{\mathbf{y}}
\newcommand{\bz}{\mathbf{z}}
\newcommand{\hx}{\mathbf{ \hat{x}}}
\newcommand{\hy}{\mathbf{ \hat{y}}}
\newcommand{\hz}{\mathbf{ \hat{z}}}
\newcommand{\hg}{\mathbf{ \hat{g}}}

\def\half{\frac{1}{2}}
\def\quart{\frac{1}{4}}
\def\ud{{\rm d}}
\def\eps{\epsilon}

\title{Irreversible mixing by unstable periodic orbits in {buoyancy dominated} stratified turbulence}

\author{Dan Lucas\aff{1}  \corresp{\email{dl549@cam.ac.uk}} \& C. P. Caulfield\aff{2,3} }
\affiliation{\aff{1}{School of Computing and Mathematics, Keele University, Staffordshire, ST5 5BG}\aff{2}{Department of Applied Mathematics \& Theoretical Physics, University of Cambridge, Centre for Mathematical Sciences, Wilberforce Road, Cambridge, CB3 0WA, UK}
\aff{3}{BP Institute, University of Cambridge, Madingley Rise, Madingley Road, Cambridge, CB3 0EZ, UK}}

\maketitle
\begin{abstract}
We consider turbulence driven by a large-scale horizontal shear in  Kolmogorov flow (i.e. with sinusoidal body forcing) and a background linear stable stratification with buoyancy frequency $N_B^2$ imposed in the third, vertical direction in a fluid with kinematic viscosity $\nu$. This flow is known to be organised into layers by nonlinear unstable steady states, which incline the background shear in the vertical and can be demonstrated to be the finite-amplitude saturation of a sequence of instabilities, originally from the laminar state. Here, we investigate the next order of motions in this system, i.e. the \emph{time-dependent} mechanisms by which the density field is irreversibly mixed. This investigation is achieved using `recurrent flow analysis'.  We identify (unstable) periodic orbits, which are embedded in the turbulent attractor, and use these orbits as  proxies for the chaotic flow. 
We find that the time average of
an appropriate measure of the `mixing efficiency' of the flow $\mathscr{E}= \chi/(\chi + \mathcal{D})$ ($\mathcal{D}$ is 
the volume-averaged kinetic energy dissipation rate and $\chi$ is the volume-averaged density variance dissipation rate) varies
non-monotonically with the time-averaged buoyancy Reynolds numbers $\overline{Re}_B= \overline{\mathcal{D}}/(\nu N_B^2)$, and 
is bounded above by $1/6$, 
consistently with the classical model 
of Osborn (1980). 
There are qualitatively different physical properties  between the unstable orbits that have lower irreversible mixing efficiency  at  low $\overline{Re}_B \sim O(1)$ and those with nearly optimal $\mathscr{E} \lesssim 1/6$ at intermediate $\overline{Re}_B \sim 10$. The weaker orbits, inevitably embedded in more strongly stratified
flow, are characterised by straining or `scouring' motions, while
the more efficient orbits have clear overturning dynamics in more weakly stratified, and apparently
shear-unstable flow.

 \end{abstract}


\section{Introduction}

A fundamental outstanding problem in stratified turbulence relates to the mixing of the stratifying agent, e.g. heat or salinity. 
Moreover such flows typically exhibit strong anisotropy with vertical motions being suppressed due to gravitational forces, and layerwise motions commonly being observed. The route by which the flow field rearranges the density field into well-defined mixed regions or `layers' separated
by sharp gradients or `interfaces' as well as the  layer/interface structure's robustness are subjects of continuing research. 
A widely utilised characterisation of mixing comes in the form of some appropriate measure of the  `mixing efficiency', often defined as the ratio of irreversible potential energy increase relative to the irreversible kinetic energy loss
 \citep{Peltier:2003gt}. One reason for this interest in mixing efficiency stems from the effort to devise accurate parameterisations of diapycnal diffusivity for use in ocean models,
and there is growing evidence that such parameterisations
should take into account variation of the efficiency with control parameters
\citep{HSalehipour:2016jr,mashayekgrl2017}.

Here our approach to explore mixing in a turbulent stratified (shear-driven) flow is  somewhat unconventional; rather than a statistical examination of increasingly large simulations or high-fidelity experimental/field data, we consider flows from the perspective of a high dimensional dynamical system and look for representative \emph{unstable} solutions embedded within the stratified turbulence. Such an approach has seen great success in clarifying the behaviour of transitional wall-bounded shear flows \citep{Kawahara:2001ft,2010PhST..142a4007C,Kawahara:2012iu} as well as the sustaining mechanisms exhibited by stationary turbulence \citep{vanVeen:2006fm,Chandler:2013fi,Lucas:2017fz}.
Recently \cite{Lucas:2017wc}, (henceforth LCK17) have shown how such an approach can advance our understanding of layer formation by locating nonlinear 
layered
steady states about which the turbulence organises.
Given this success,  we are motivated to investigate  whether  the next order of motions can be identified, those time-dependent simple invariant manifolds, i.e. periodic orbits, which capture some salient signature of the processes by which buoyancy is mixed. In particular, we are interested in  how `efficient' (defined in a precise fashion below)  such mixing is, and how such processes vary with  control parameters.

To address these issues, the paper is organised as follows. Section 2 contains the formulation and discussion of the methods employed, while section 3 presents a set of preliminary and motivational direct numerical simulations (DNS) with a discussion of their mixing properties. Section 4 shows results from the recurrent flow analysis together with a discussion of the processes exhibited by the periodic orbits discovered,
and finally section 5 presents our conclusions. 


\section{Formulation}
We begin by considering the following version of the non-dimensionalised, monochromatic body-forced, incompressible, Boussinesq equations 
\begin{align}
\frac{\partial \bm u}{\partial t}& + \bm u\cdot\nabla\bm u +\nabla p 
= \frac{1}{Re} \Delta \bm u + \sin( n y)\hx - B \rho \hz\label{NSu},\\ 
\frac{\partial \rho}{\partial t}& + \bm u\cdot\nabla\rho = w + \frac{1}{Re Pr}\Delta \rho, \qquad \qquad
\nabla\cdot \bm u =0
\end{align}
where we define the (external) Reynolds number $Re$, the bulk stratification parameter $B$ and the 
Prandtl number $Pr$ as
\begin{equation}
Re := \frac{\sqrt{\lambda}}{\nu}\left(\frac{L_y}{2\pi}\right)^{3/2}, \quad B:= \frac{g\beta L_y^2}{\rho_0 \lambda 4 \pi^2}, \qquad Pr = \frac{\nu}{\kappa}.
\end{equation}
Here, $\bm u(x,y,z,t) = u \hx+v \hy+w \hz$ is the three-dimensional
velocity field, $p$ is the pressure and the density is decomposed into $\rho_{tot}=\rho_0+\rho_B(z) +\rho(\bm x,t)$, i.e as the sum of a Boussinesq reference density, a constant linear background stratification and a fully varying disturbance density. We have  non-dimensionalised using the  characteristic length scale $L_y/2\pi$, characteristic time scale $\sqrt{L_y/2\pi\lambda}$ and density gradient scale $\beta=\ud\rho^*_B/\ud z.$ (see LCK17 for details). Furthermore, 
 $n$ is the forcing wavenumber, $\lambda$ is the forcing
amplitude, $\nu$ is the kinematic viscosity, $\kappa$ is the molecular diffusivity. We impose periodic boundary conditions in all directions and  solve over the cuboid $[0,2\pi/\alpha]\times[0,2\pi]^2$ where $\alpha=L_y/L_x$ defines the horizontal aspect ratio of the domain. Vorticity $\bm \omega = \nabla \times \bm u$ is used as the prognostic variable and DNS are performed using the fully dealiased (two-thirds rule) pseudospectral method with mixed fourth order Runge-Kutta and Crank-Nicolson timestepping implemented in CUDA to run on GPU cards.  We initialise the velocity field's Fourier components with uniform amplitudes and randomised phases in the range $2.5 \leq |\bm k | \leq 9.5$ such that the total enstrophy $\langle |\bm \omega |^2\rangle_V = 1$ and  $\rho'=0$ initially.  Throughout,  $\langle (\cdot)  \rangle_V:=\alpha \int \! \! \int \! \! \int (\cdot)\,dxdydz/(2\pi)^3$ denotes a volume average, $\langle (\cdot)  \rangle_h:=\alpha \int \! \! \int (\cdot)\,dxdy/(2\pi)^2$ denotes a horizontal average and $\langle (\cdot)  \rangle_v:= \int \! (\cdot)\,dz/(2\pi)$ denotes a vertical average.


We define the diagnostics involved in the energetic budgets as
\begin{align} 
\mathcal{K} &=\frac{1}{2}\langle|\bm u|^2\rangle_V, \quad 
\mathcal{P} =\frac{B}{2}\langle \rho^2\rangle_V, \quad
\mathcal{I} = \langle \bm u \cdot \bm f\rangle_V = \langle  u  \sin(ny) \rangle_V, \label{eq:diag1}\\
\mathcal{B} &= \langle \bm u \cdot B \rho\hz \rangle_V= B\langle w\rho \rangle_V, \quad
\mathcal{D} = \frac{1}{Re}\langle|\nabla\bm u|^2\rangle_V, \quad \chi = \frac{B}{PrRe}\langle|\nabla \rho|^2\rangle_V,
\label{eq:diag2}
\end{align}
where $  \frac{\ud \mathcal{K}}{\ud t} = \mathcal{I} - \mathcal{B}-\mathcal{D}, \quad 
\frac{\ud \mathcal{P}}{\ud t} =  \mathcal{B}-{\chi}$ and
$\mathcal{K}$ is the total kinetic energy density, $\mathcal{P}$ the density variance, $\mathcal{I}$ is  the energy input by the forcing,  $\mathcal{B}$ is the buoyancy flux,  $\mathcal{D}$ is the viscous dissipation rate, and $\chi$ the density variance dissipation rate.  { We fix $\alpha=0.5$ to avoid subcritical transition, $Pr=1$ for numerical efficiency and $n=1$, i.e. the flow is forced with $\sin(y)\hx,$ to  mimic closely other large scale shear profiles previously studied (e.g. stratified Taylor-Couette flow \citep{woodsjfm2010} and vertically sheared Kolmogorov flow \citep{Garaud:2015ce})}  and denote time averages with overbars, i.e.
$\overline{(\cdot )} = [\int_0^T (\cdot) \ud t]/T$ where $T$ is normally the full simulation time.
\section{Irreversible mixing in the direct numerical simulations}

We begin by characterising the mixing in this system by conducting a set of simulations across a range  of $Re$ and $B.$ The pertinent single point diagnostics are presented in table \ref{tab:DNS}. In particular we examine the irreversible mixing efficiency which we, following \cite{Salehipour:2015bt} and \cite{Maffioli:2016bw}, define in terms of the dissipation rates:
\beq
\mathscr{E}(t)=\frac{\chi}{\chi+\mathcal{D}}.
\eeq

The dependence of $\mathscr{E}$ on  typical flow parameters is a crucial ongoing problem in stratified turbulence, and a source of some controversy. As discussed by \cite{matergrl2014}, the flow of a turbulent, shear-driven
stratified fluid has several competing time scales, and it is natural
to attempt to parameterise $\mathscr{E}$ 
in terms of parameters quantifying
the relative importance of these time scales. Using shear-instability simulations
and comparison with observations, \cite{HSalehipour:2016jr} developed a parameterisation
using two parameters: an appropriately defined `buoyancy Reynolds number', effectively a ratio
of the time scale of the stratification
to the time scale of the turbulence; and an appropriately defined Richardson number, effectively a ratio 
of the square of the time scale of the vertical shear (assumed to be the dominant driver
of the mixing) to the time scale of the stratification. {In order to determine how the mixing efficiency varies with these dimensionless numbers in this flow, we define the buoyancy Reynolds number $Re_B(t)$ and the (local) gradient Richardson number
$Ri_G(\bm x, t)$ as (with the usual caveat that comparing  different studies is difficult if the key parameters are defined differently)}
\begin{align} Re_B(t)&=\frac{\mathcal{D} Re}{B}, \qquad
 Ri_G(\bm x,t) = \frac{-B\frac{\partial\rho_{tot}}{\partial z}}{\left(\frac{\partial u_h}{\partial z}\right)^2},
\label{eq:RiG_ReB} 
\end{align}
where $(\partial u_h/\partial z)^2=(\partial u/\partial z)^2+(\partial v/\partial z)^2.$
 $Ri_G$ is a pointwise quantity which characterises the relative stability of the flow to overturning shear instabilities, and developing an appropriate overall average for this quantity to characterise a
particular flow should be performed with care, not least because locations of low vertical shear significantly skew the distributions of $Ri_G.$ The appropriate averaging used here is thus 
\beq
{Ri}_B(t)=  \left \langle \frac{\left \langle-B\frac{\partial\rho_{tot}}{\partial z}\right \rangle_h}{\left \langle \left(\frac{\partial u_h}{\partial z}\right)^2\right \rangle_h} \right \rangle_{z}. \label{eq:RiG}
\eeq

There are two important points to appreciate about these definitions. First, we have  chosen to define
$Ri_B$ with  the total gradients of density while $Re_B$ only depends explicitly on the background $N_B.$
Second, although in principle these are independent nondimensional parameters, it remains
to be established that they are in this flow, as there
exist situations where they are strongly correlated (see e.g. \cite{Zhou:2017dd}).

\begin{figure}
\includegraphics[width=\textwidth]{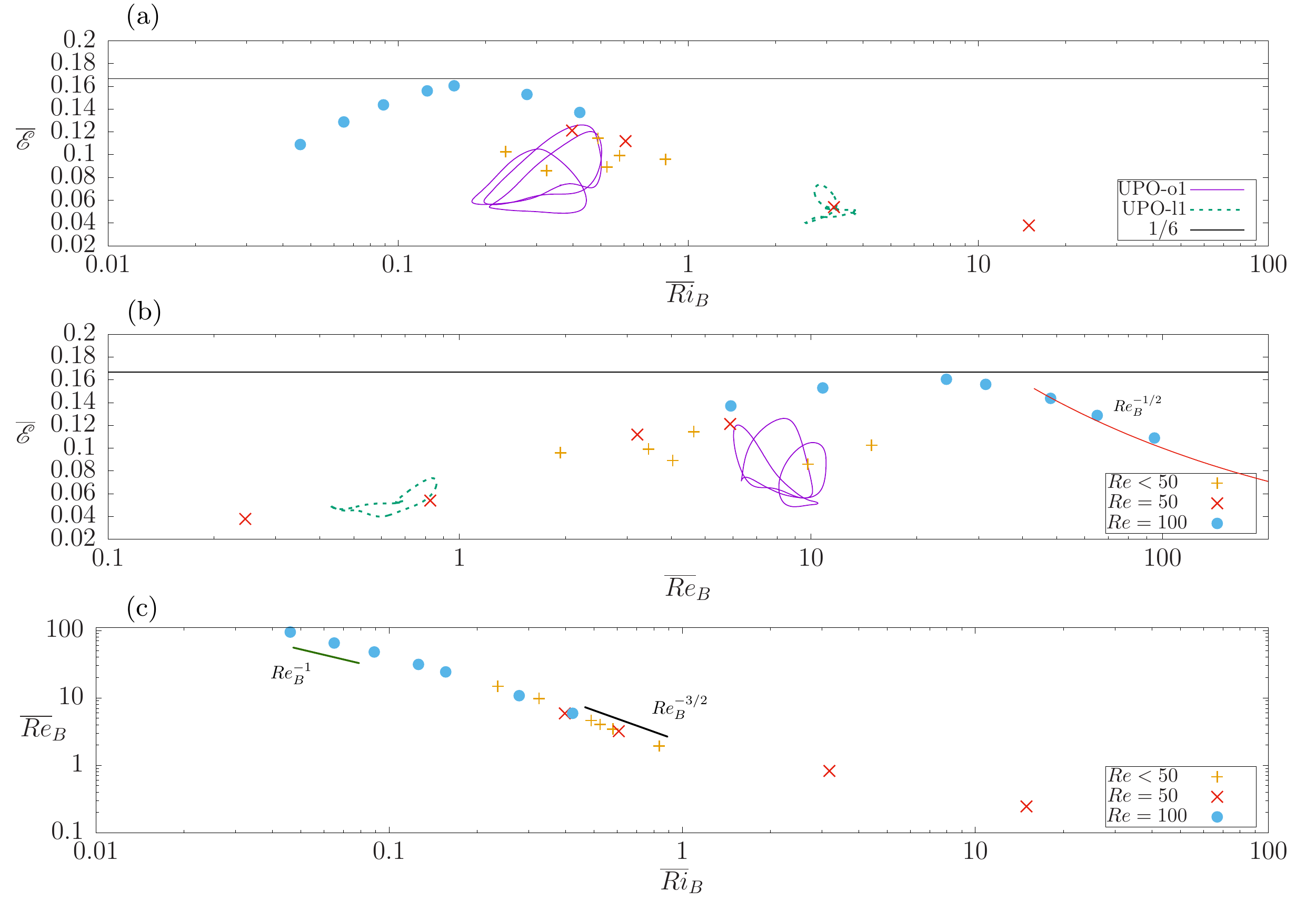}
\caption{The variation of average mixing efficiency $\overline{\mathscr{E}}$ with: (a) average bulk Richardson number $\overline{Ri}_B$  and (b) average buoyancy Reynolds number $\overline{Re}_B$. The horizontal line is $\overline{\mathscr{E}}=1/6$, consistent with 
the upper bound of \cite{Osborn:1980jj}. The closed loops mark the projections onto the time-dependent $(Ri_B,\mathscr{E})$ and 
$(Re_B,\mathscr{E})$
planes of the two unstable relative periodic orbits UPO-o1 (solid line) and UPO-l1 (dashed line) discussed in section \ref{sec:UPOs}. 
(c) shows the variation of $\overline{Re}_B$  with $\overline{Ri}_B$. 
Different symbols mark different values of $Re$. The red curve in (b) corresponds to $\overline{\mathscr{E}}=\overline{Re}_B^{-1/2}$, the large $Re_B$ behaviour suggested by \cite{SHIH:2005bx}. }
\label{fig:mix_ReB}
\end{figure}

In figure \ref{fig:mix_ReB}(a) and (b)  we plot the time-averaged mixing efficiency $\overline{\mathscr{E}}$ against the time-averaged parameters $\overline{Ri}_B$  and $\overline{Re}_B$. 
We observe non-monotonic dependence of $\overline{\mathscr{E}}$ on both parameters, 
in at least qualitative agreement with the experimental analysis of \cite{linden1979} 
and numerical simulations of \cite{SHIH:2005bx}. Indeed, $\mathscr{E}$ appears to 
saturate near the critical value of $1/6$, (marked with a horizontal line) equivalent to the upper bound on the turbulent  flux coefficient $\Gamma \simeq \mathscr{E}/(1-\mathscr{E}) \leq 0.2$ proposed by \cite{Osborn:1980jj} and then decreases for $\overline{Re}_B\gtrsim30,$ with a dependence 
not entirely unlike the $\overline{\mathscr{E}} \propto \overline{Re}_B^{-1/2}$ observed
by \cite{SHIH:2005bx} in what they referred to as the `energetic' regime (see
\cite{Ivey:2008hm} for a more detailed discussion). 

However, although these similarities are intriguing, there are still significant differences.
Firstly, we do not observe a `transitional' plateau of approximately constant mixing efficiency at intermediate
$\overline{Re}_B$. Indeed, the  qualitative structure of the mixing efficiency curve is more reminiscent of the Pad\'e approximant
proposed by \cite{mashayekgrl2017} using (vertical) shear-instability numerical data as well as observations, 
where $\Gamma \propto Re_B^{1/2}$ and $Re_B^{-1/2}$ for small and large $Re_B$ respectively.
Second, and once again reminiscent of the approach proposed by \cite{mashayekgrl2017} to parameterise
mixing in terms of $\overline{Re}_B$ alone, 
the two parameters  $\overline{Ri}_B$  and $\overline{Re}_B$ are actually closely 
correlated, and our simulations actually trace out
a (monotonic) curve in $\overline{Ri}_B-\overline{Re}_B$ space, as shown in figure \ref{fig:mix_ReB}(c),
independently of the externally imposed parameters $B$ and $Re$.
The low mixing efficiency observed at high $\overline{Ri}_B$ can thus be understood as being
entirely associated with weaker turbulence (smaller $\overline{Re}_B$)  and the maximum
mixing efficiency occurs at a sweet spot of sufficiently
vigorous turbulence at sufficiently high stratification, analogously
to the results of \cite{Zhou:2017dd}. We stress that
we are not claiming that this clear correlation between $\overline{Ri}_B$ and $\overline{Re}_B$ is 
generic, (see for example the discussion in \cite{scottijpo2016}) just that it 
occurs for this flow.

It is apparent 
that at small  $\overline{Ri}_B$,  $\overline{Ri}_B \propto \overline{Re}_B^{-1}$. This is unsurprising, 
as the turbulence and shear are largely unaffected by such weak 
stratification, and the variation of both parameters with $B$ completely dominates.
As $\overline{Ri}_B$ increases beyond the value associated with the most efficient
mixing however, the turbulence and shear do indeed become suppressed by the 
strengthening stratification, and the power law dependence
steepens 
so $\overline{Ri}_B \propto \overline{Re}_B^{-3/2}$. 
There is undoubtedly also an increasing significance of viscosity, which
becomes dominant for the largest values of $\overline{Ri}_B > 1$, where $\overline{Re}_B$ drops below one, and 
the turbulence is essentially completely suppressed,
with very
inefficient mixing. 
However, it is always important to remember that both $\overline{Ri}_B$ and 
$\overline{Re}_B$  are global measures averaged in both space and time.
Both $Ri_B$ and $Re_B$ are functions of time, and
there could also be 
spatial 
variation of $Ri_G$ within the flow domain.
Such variation proves to be
a key part of the behaviour of the periodic orbits which we identify,
and hence of the mixing occuring in these flows.  

\begin{table} \setlength{\tabcolsep}{8pt}
\begin{center}
\begin{tabular}{cccccccccc}
\#&$Re$& $B$ & $\overline{Ri}_B$ &  $\eta k_{\mathrm{max}}$& $\overline{\mathscr{E}}$& $\overline{Re}_B$  & $l_o$  & $Re_\lambda$ \\
\hline
A1&20&1	&	0.325&	2.7	&	0.0859	&	9.8	&	0.695		&	31.3 \\
A2&30&1	&	0.235&	2.0	&	0.1026	&	14.9	&	0.698		&	51.2 \\
A3&30&5	&	0.525&	1.8	&	0.0892	&	4.04	&	0.244		&	25.5 \\
A4&40&5	&	0.489&	1.5	&	0.114	&	4.64	&	0.226		&	52.1 \\
A5&40&7.5	&	0.58&	1.5	&	0.099	&	3.45	&	0.177		&	31.6 \\
A6&40&10 &	0.835&	1.6	&	0.096	&	1.94	&	0.124		&	23.6 \\[6pt]
B1&50&5	  &	0.398&	1.3	&	0.121	&	5.89	&	0.228		&	44.7 \\
B2&50&10  &	0.608&	1.3	&	0.112	&	3.21	&	0.141		&	44.1 \\
B3&50&50  &	3.18&	1.2	&	0.054	&	0.83	&	0.048		&	31.3 \\
B4&50&100&	14.95&	1.4	&	0.038	&	0.25	&	0.022		&	45.9\\[6pt]
C1&100&0.5 &	0.046&	1.6	&	0.109&	94.82	&	1.15			&	98.5 \\
C2&100&0.75& 0.065&	1.6	&	0.129&	65.19	&	0.86			&	92.5 \\
C3&100&1.0 &	0.089&	1.6	&	0.144&	48.03	&	0.69			&	94.1 \\
C4&100&1.5 &	0.126&	1.6	&	0.156&	31.41	&	0.50			&	91.2 \\
C5&100&2.0 &	0.156&	1.6	&	0.160&	24.3		&	0.41				&	88.6 \\
C6&100&5.0 &	0.278&	1.5	&	0.153&	10.8		&	0.22				&	76.2 \\
C7&100&10.0 &0.423&	1.5	&	0.137&	5.91		&	0.14			&	62.9 \\

\end{tabular}
\end{center}
\caption{\label{tab:DNS} Imposed parameters and diagnostic outputs for the three groups of simulations.  The 
nondimensional Kolmogorov microscale
is given by $\eta=Re^{-3/4} \overline{\mathcal{D}}^{-1/4}$
and is scaled by the  maximum wavenumber allowable $k_{\mathrm{max}}=N/3$. 
Groups A and B have resolution $128\times64^2$ and group C has $256\times128^2$ to retain spectral convergence. The nondimensional Ozmidov length scale $l_O={\overline{\mathcal{D}}}^{1/2}/{B^{\frac{3}{4}}}$, the  average buoyancy Reynolds number $\overline{Re}_B=\overline{\mathcal{D}} Re/B,$ the average bulk Richardson number $\overline{Ri}_B$ 
and the Taylor microscale Reynolds number $Re_\lambda=\mathcal{K}_{\mathrm{turb}}\sqrt{{10Re}/{\overline{\mathcal{D}}}}$ 
are also listed. Here we define $\mathcal{K}_{turb} = (1/2)\langle(\bm u - \overline{\bm u})^2\rangle_V$ and overbars are always time averages over the full $T=1000$ window.  
}
\end{table}

\section{Recurrent flow analysis}\label{sec:UPOs}

In order to determine what processes underpin the turbulence in our simulations, we use   `recurrent flow analysis'. The general approach is as follows. First a simulation is conducted during which near recurrences are located in the chaotic trajectory. This is achieved by storing a historical record of state vectors and periodically checking the history against the current state vector. When an appropriately `near repeat' is identified, it is stored for a later convergence attempt with a Newton-GMRES-hookstep algorithm. For more  details on this approach see \cite{Chandler:2013fi,Lucas:2015gt}.
Here, we are interested in finding representative unstable periodic orbits (UPOs) with qualitatively
different mixing properties 
in a broad range of flows, rather than identifying
a large number of UPOs for a specific set of parameters. 
Therefore, we have identified five orbits in two broad classes: one class 
of two orbits found in the relatively weakly stratified  simulation A1 with $Re=20$ and $B=1$,
and the other class of three orbits in the more strongly stratified simulation B3 with $Re=50$ and
$B=50$. The first class has $\overline{Re}_B \sim O(10)$, and 
we label it as class `o', for `overturning', while
the second class has $\overline{Re}_B \sim O(1)$, and we label it as class `l' for layered. We plot projections 
onto the time-dependent $(Ri_B,\mathscr{E})$ and 
$(Re_B,\mathscr{E})$
planes of the 
trajectories of characteristic  relative periodic orbits UPO-o1 (solid line) and UPO-l1 (dashed line)
on figures \ref{fig:mix_ReB}(a) and (b), demonstrating that these two classes do indeed 
have qualitatively different mixing properties, and also that these time-dependent properties are consistent
with the time-averaged properties of the `full' DNS.

\begin{table}
\begin{center}
\begin{tabular}{llllcccccccccc}\label{big_table}
No.  & DNS & $Re$&$ B$&  $T_p$  &   $s_x$  &  $s_z$  &  $m_y$  &  $\Lambda^{-1}$  &  $N_\lambda$  & $\overline{Re}_B$&$\max(\mathscr{E})$ & $\overline{Ri}_B$ & $\overline{\mathscr{E}}$\\
\hline 
& & & & & & & &\\
UPO-o1 & A1 &20 & 1 &21.28 & 5.98 &  0.35 & 0 & 1.65 &25 & 8.5 & 0.13 & 0.333 & 0.078  \\ %
UPO-o2  & A1 &20 & 1 & 7.51 & 7.81 & -0.06 & 0 & 2.30 &27 & 8.9 & 0.097& 0.265 & 0.067 \\
UPO-l1 & B3 & 50& 50& 8.75  & 0.016 & 0  & 0 & 0.211 & 7 & 0.62 & 0.07 & 3.17 & 0.052 \\ %
UPO-l2 & B3 &  50& 50&8.74 & 0 &  0 & 0 &  0.207& 6 & 0.63 & 0.058  & 3.13 & 0.048 \\ %
UPO-l3 & B3 &  50& 50&9.21 & 0 &  0.007 & 0 & 0.116 & 6& 0.57 & 0.051 & 4.64 & 0.045  \\ %

\hline 
\end{tabular}
\caption{ Table cataloguing the unstable recurrent flows showing the DNS from which they come (table 1), period $T_p$, relative shifts in $x$ and $z$ directions- $s_x$, $s_z$ discrete shift/reflect in $y$ $m_y,$ inverse stability coefficient ($\Lambda^{-1}$, see \cite{Chandler:2013fi}), the number of unstable directions $N_\lambda$,  $\overline{Re}_B,$ maximum and time averaged mixing efficiency $\mathscr{E}$ and time averaged bulk Richardson number $\overline{Ri}_B$.\label{tab:UPOs}}\end{center}
\end{table}

\begin{figure}
\hspace{-5mm}\scalebox{0.43}{\input{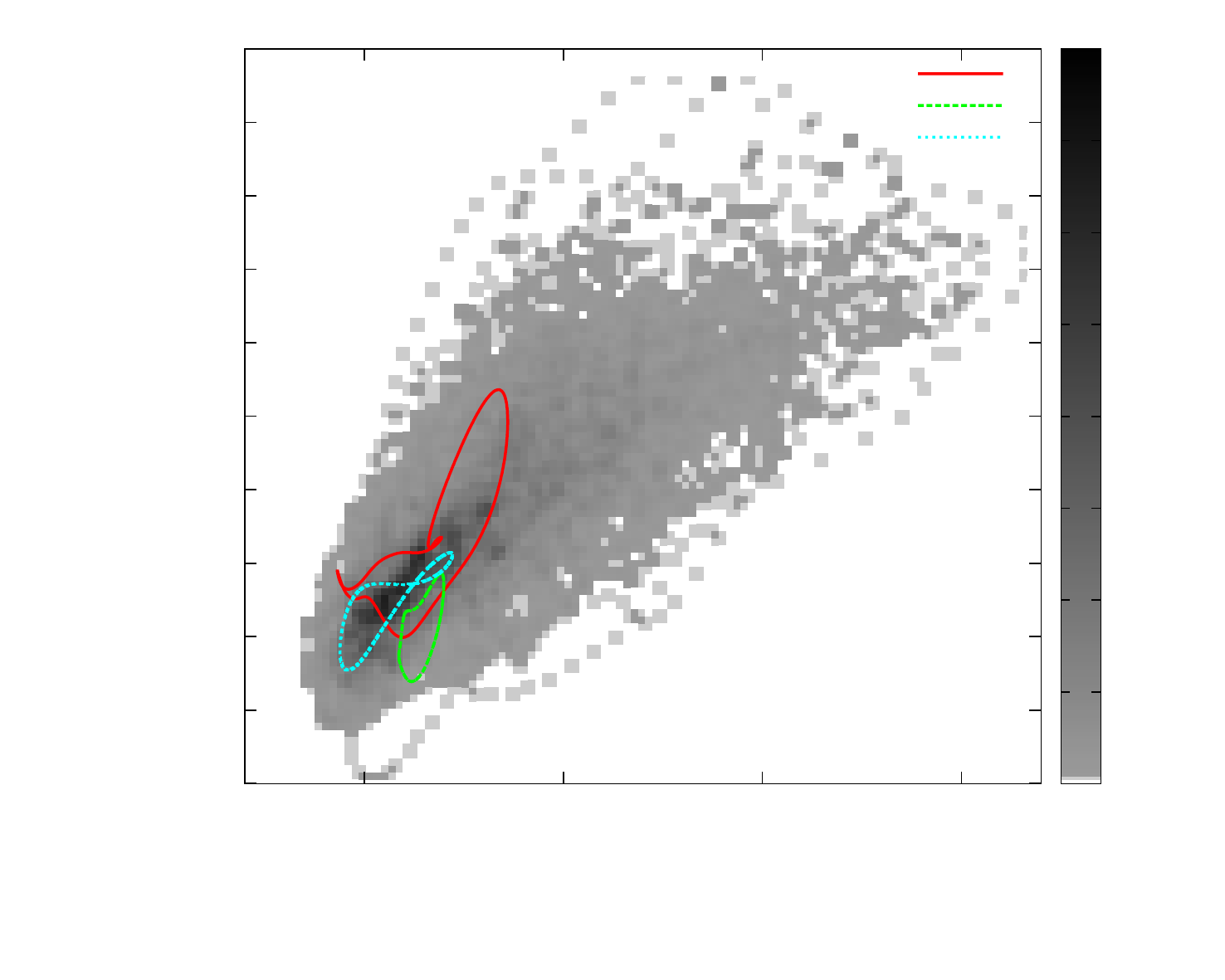}}\hspace{5mm}
\scalebox{0.43}{\input{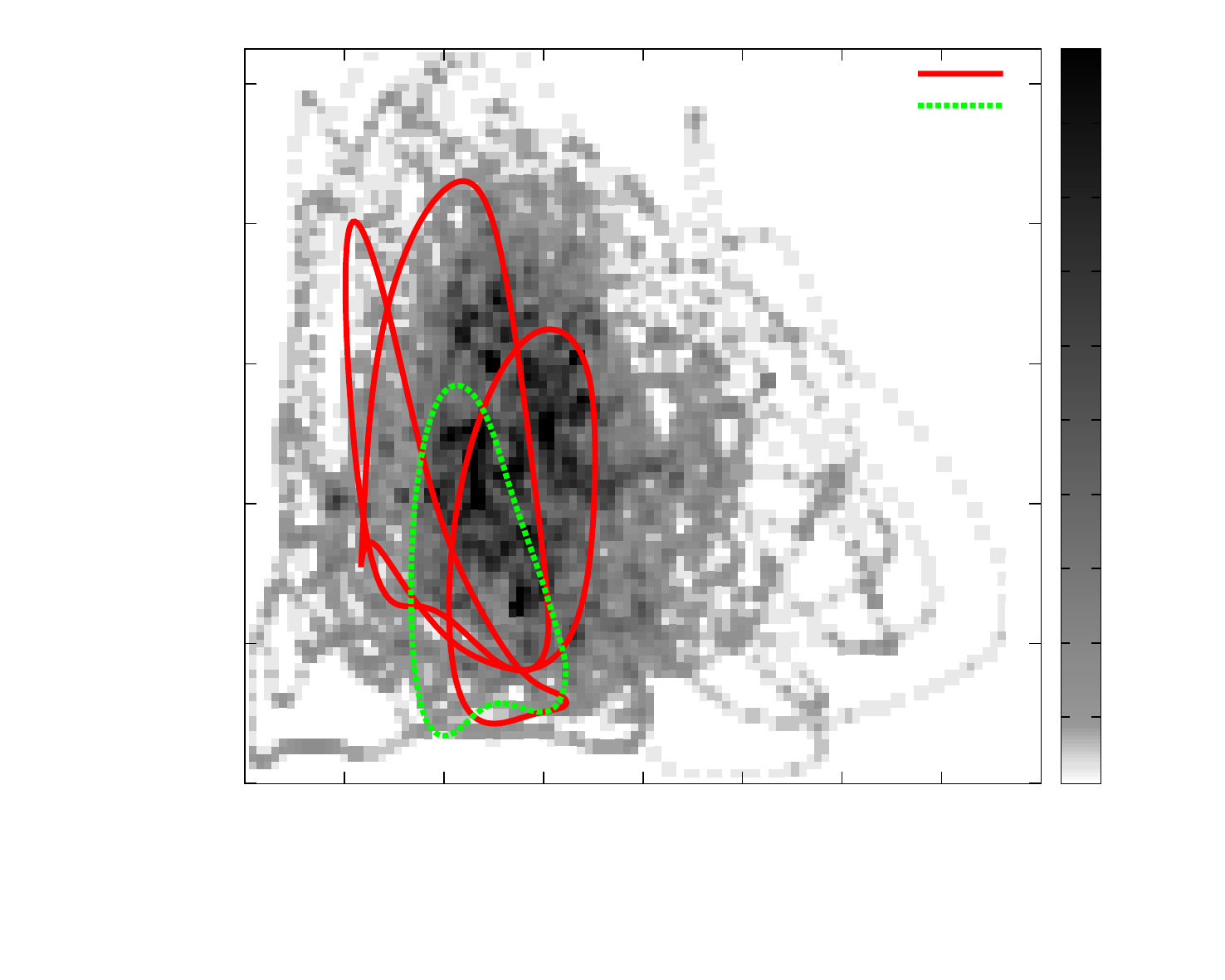}}
\caption{Projection onto the ($Re_B,\mathscr{E}$) plane of the UPOs and the p.d.f. of the direct numerical simulations rendered in greyscale, with darker shades denoting that the turbulent trajectory spends more time there for: (a) class `o' UPOs from simulation A1; and (b) class `l' UPOs from simulation B3.
} \label{fig:eff_ReB}
\end{figure}

Table \ref{tab:UPOs} lists properties of the converged orbits,  including period, relative shifts due to the continuous symmetries in $x$ and $z,$ stability and some diagnostic averages. {By comparison with the averages of the DNS in table \ref{tab:DNS}, it is clear the UPOs are reproducing the bulk time-averages quite well.} The 
projection of the trajectories onto the $(Re_B, \mathcal{E})$ plane of the 
five converged orbits are plotted in 
figure \ref{fig:eff_ReB}, with greyscale colours representing the probability density function (p.d.f.) of the turbulent DNS. {Notice that the darker colours represent regions where the turbulent trajectory spends more time, and lighter colours represent less frequent excursions to that part of phase space. For simulation B3,   the `l' class of converged UPOs sit across the darkest region of the p.d.f. where the turbulence spends most of its time but miss the higher efficiency, intermittent turbulent bursts that the DNS exhibits. Missing extreme events is a known failing of the recurrent flow analysis in some circumstances and is a topic of ongoing research, see \cite{Lucas:2015gt}. However,
for  the `o' class in simulation A1, the orbit UPO-o1 (in particular) appears to span the turbulent attractor well in this projection, missing only some very rare excursions to high $Re_B.$} As shown in table \ref{tab:UPOs},  the `o' class orbits are considerably more unstable than the orbits in `l' class.  

\begin{figure}
\hspace{5mm}
\includegraphics[width=\textwidth]{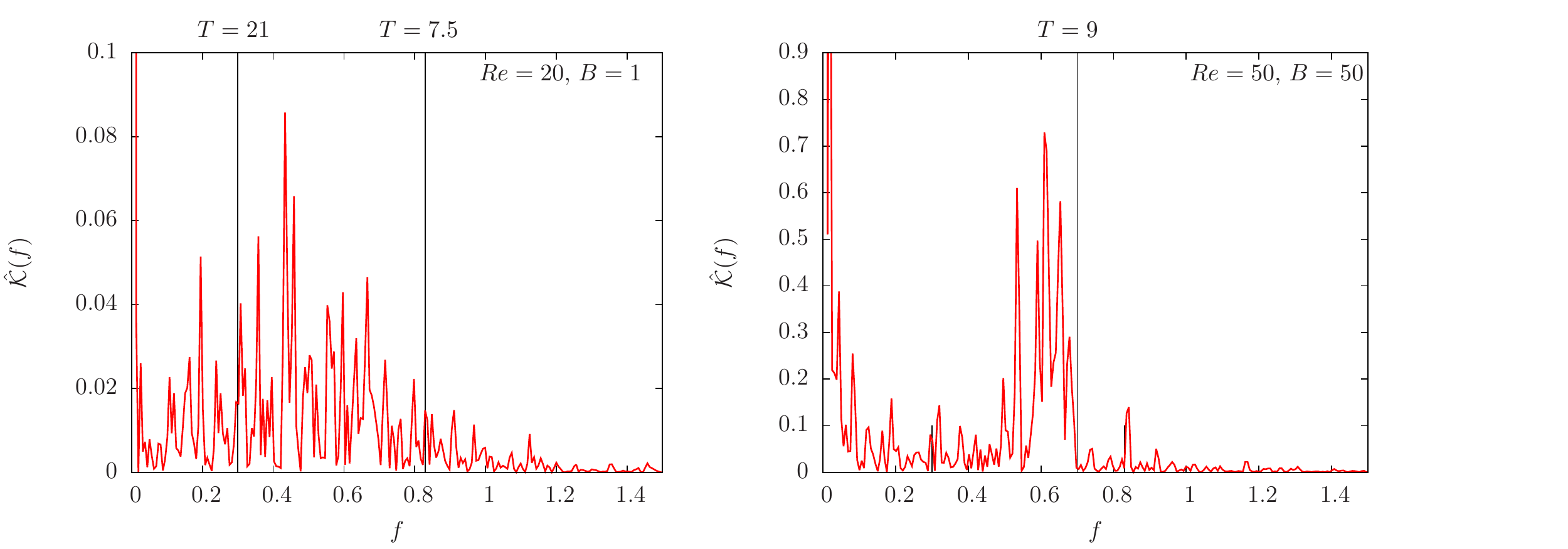}
\caption{Power spectra for the kinetic energy $\hat{\mathcal{K}}(f)=\int_T\mathcal{K}(t) \mathrm{e}^{\mathrm{i}ft}\ud t$ for the DNS signals from simulation A1 (left) and simulation B3 (right) with vertical lines showing approximately the periods of the UPOs extracted.} \label{fig:pspec}
\end{figure}

{In order to compare how the timescales of the UPOs relate to those exhibited by the turbulence,  we plot in figure \ref{fig:pspec} the power spectra of the kinetic energy for the two DNS signals from which the UPOs have been extracted (i.e. A1 and B3). Vertical lines show that the the periods of the orbits are approximately coincident with observed frequencies. For simulation A1  the largest non-zero peak in the frequency spectrum occurs at approximately half the fundamental frequency observed in the UPOs, i.e. $f\approx 2\pi/15 =0.42$ with the fundamental period around $7.5$, UPO-o1 having just under three such periods. For simulation B3 the periods of the orbits are on the edge of the main distribution of frequencies near $f=0.6.$ This is reasonable evidence that the timescales of the UPOs are representative of the typical timescales of the turbulence. We can also consider the buoyancy frequency and its influence on these cases. For simulation A1 where $B=1$, the background buoyancy frequency $N_B=\sqrt(B)=1$ so the timescale  or `buoyancy period' is $T_N=2\pi$ which is not far from the ``fundamental period'' we have established from the UPOs and spectrum. For B3,  the buoyancy period $T_N=2\pi/\sqrt(50)\approx 0.9$ which is much faster than the timescales found in the spectrum. We find some signature of this frequency later in this section.} Henceforth, we will focus attention on UPO-o1 and UPO-l1 since, having larger amplitude, they represent better proxies for the turbulence, although the dynamics within each class is broadly similar. 
%

\begin{figure}
\scalebox{0.38}{\input{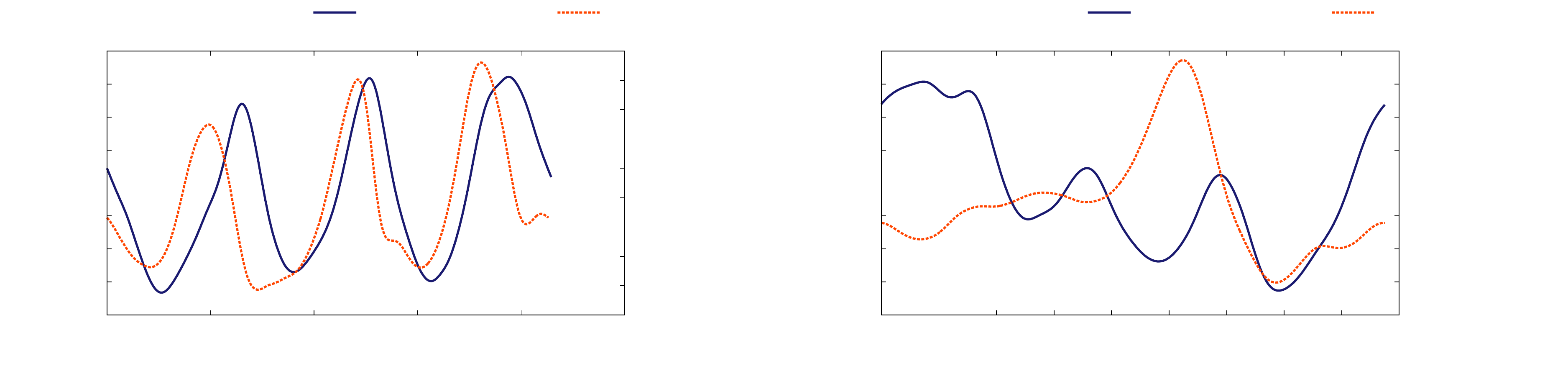}}
\caption{Time evolution of $Ri_B$ and $\mathscr{E}$ for: (a)  UPO-o1; and (b) UPO-l1. Notice the correlation between $Ri_B$ and $\mathscr{E}$ for (a) and not for (b). The times highlighted in figures \ref{fig:3D_UPO2} and \ref{fig:3D_UPO1} are marked with stars. \label{fig:RiG_eff}}
\end{figure}

\begin{figure}
\includegraphics[width=\textwidth]{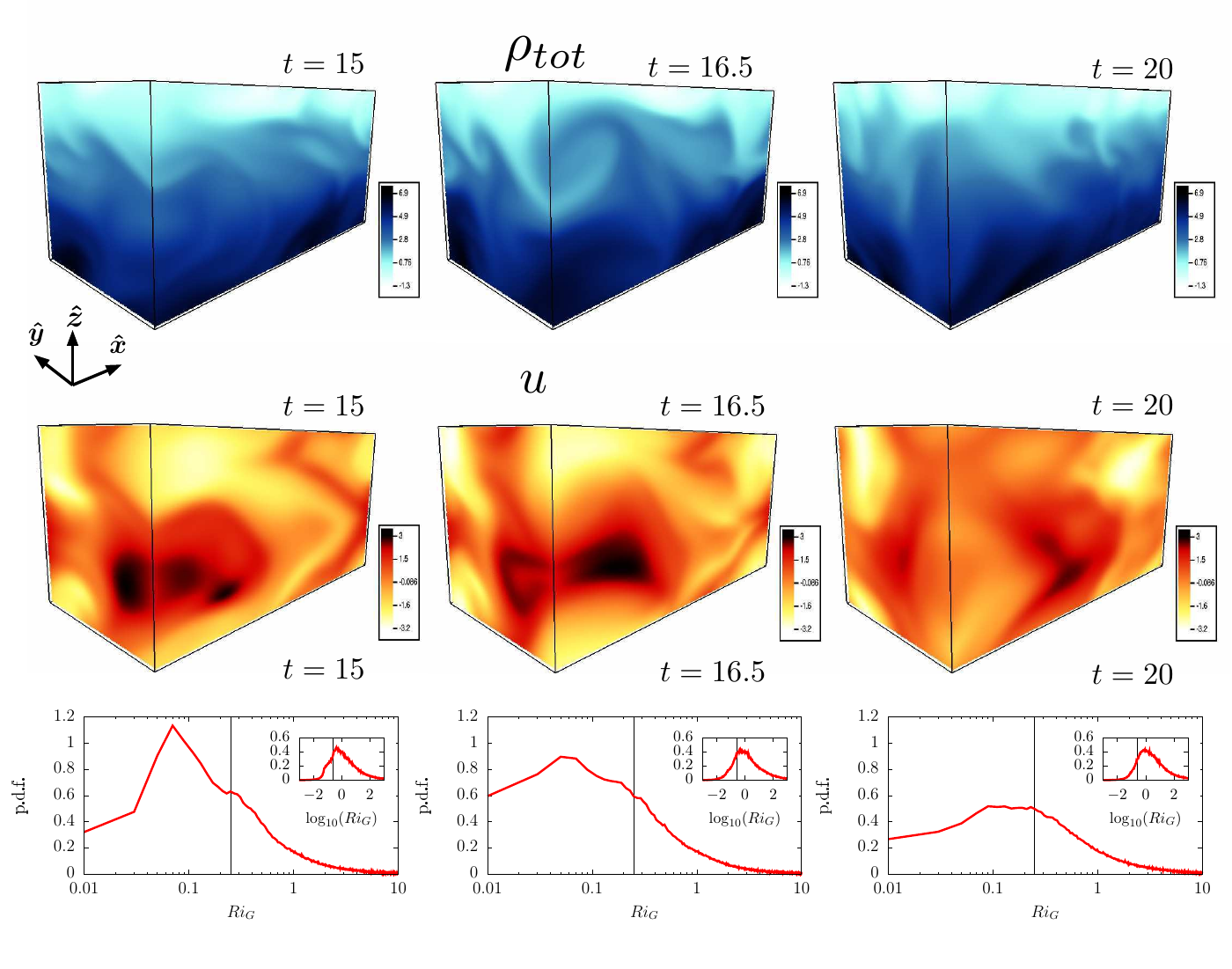}
\caption{ 
Three dimensional 
 rendering of: (top)  $\rho_{tot}=\rho-z$;  (middle)  $u$;
 and (bottom) p.d.f.s of $Ri_G$ in the entire computational domain 
at $t=15,\,16.5$ and $20$) (left to right) for UPO-o1. The p.d.f.s are computed with linear intervals of $Ri_G,$ 
while the inset shows the result when distributing the intervals on a log scale.  Vertical lines mark $Ri_G=1/4$.} \label{fig:3D_UPO2}
\end{figure}

Further evidence that the two classes are qualitatively different is shown in figure \ref{fig:RiG_eff}, 
where we plot the time dependence of $Ri_B$ and $\mathscr{E}$ for UPO-o1 and UPO-l1.
The dynamics of UPO-o1 show relatively low $Ri_B$, strongly correlated to the mixing efficiency $\mathscr{E}$. There are three distinct sub-periods, each of which shows the bulk $Ri_B$ decrease to small values (indeed
below the canonical value of $1/4$ {for the linear instability of a stratified (vertical) shear layer given by the Miles-Howard theorem and frequently invoked as a diagnostic for overturning}), followed by a maximum in mixing efficiency as the flow overturns. Such overturnings are  visualised in figure \ref{fig:3D_UPO2}, where {snapshots of total density $\rho_{tot}$ and streamwise velocity $u$ are chosen before and after the final minimum of $Ri_B$ and near the subsequent maximum of $Ri_B$ following the peak $\mathscr{E}$ as marked
on figure \ref{fig:RiG_eff}(a).}
(An animation of the time evolution of these fields is available as supplementary material.)
{The density field shows very distinct Kelvin-Helmholtz-like billows forming at $t=16.5$ where $\mathscr{E}$ and $Ri_B$ are beginning their growth phase, justifying the labelling of this UPO as being in class `o'.}
Of course, a bulk measure of the Richardson number does not capture the stability properties 
of the flow, 
and so in figure \ref{fig:3D_UPO2} we also plot the p.d.f. of $Ri_G(\bf{x})$ over the spatial domain at the same times,  with a vertical line indicating $Ri_G=1/4$. These distributions show a striking increase of the proportion of the domain with $Ri_G<1/4$ prior to the overturning mixing event. This proportion then decreases with time, but does not vanish completely. Even in the periods of maximum \emph{bulk} $Ri_B$ some regions of the domain remain with $Ri_G < 1/4$.  Apparently, it is necessary for an appreciable proportion of the domain to have a local gradient Richardson number $Ri_G < 1/4$ before overturning is observed. 

\begin{figure}
\includegraphics[width=\textwidth]{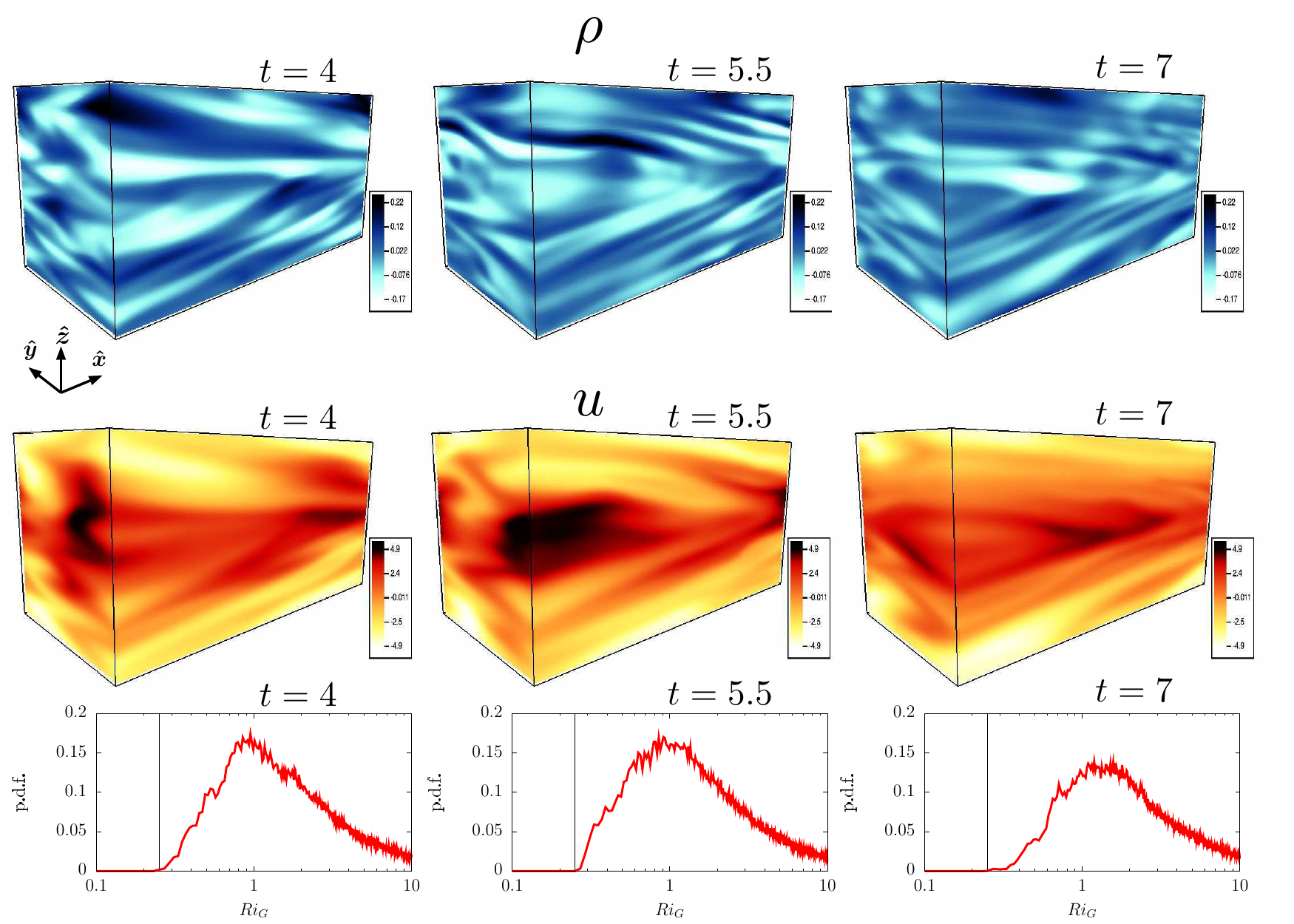}
\caption{
Three dimensional 
 rendering of: (top)  $\rho_{tot}=\rho-z$;  (middle)  $u$;
 and (bottom) p.d.f.s of $Ri_G$ in the entire computational domain 
at $t=4,\,5.5$ and $7$) (left to right) for UPO-l1. The p.d.f.s are computed with linear intervals of $Ri_G,$ 
while the inset shows the result when distributing the intervals on a log scale.  Vertical lines mark $Ri_G=1/4$.
}
\label{fig:3D_UPO1}
\end{figure}

By contrast, as is apparent in figure \ref{fig:RiG_eff}(b),  UPO-l1 has much larger overall $Ri_B$, seemingly uncorrelated to the behaviour of the markedly lower mixing efficiency $\mathscr{E}$. This is 
also apparent in the snapshots shown in figure \ref{fig:3D_UPO1} at the characteristic times
marked on figure \ref{fig:RiG_eff}(b)
(and the associated animations available as supplementary materials).
 No  
overturns occur, and  the cycle of mixing behaviour is associated with straining or scouring motions drawing the perturbation density $\rho$ into thin layers, justifying the labelling as class `l'. The instantaneous p.d.f.s of the gradient $Ri_G$ plotted in figure \ref{fig:3D_UPO1} now show that nowhere in the domain has $Ri_G<1/4,$ even near $t=5.5$ when mixing efficiency is maximised. At peak $\mathscr{E},$ $t\approx5.5,$ the snapshot of $\rho$ shows increased layers and arguably larger $\partial \rho /\partial z.$ 

\begin{figure}
\hspace{-2mm}\includegraphics[width=1.1\textwidth]{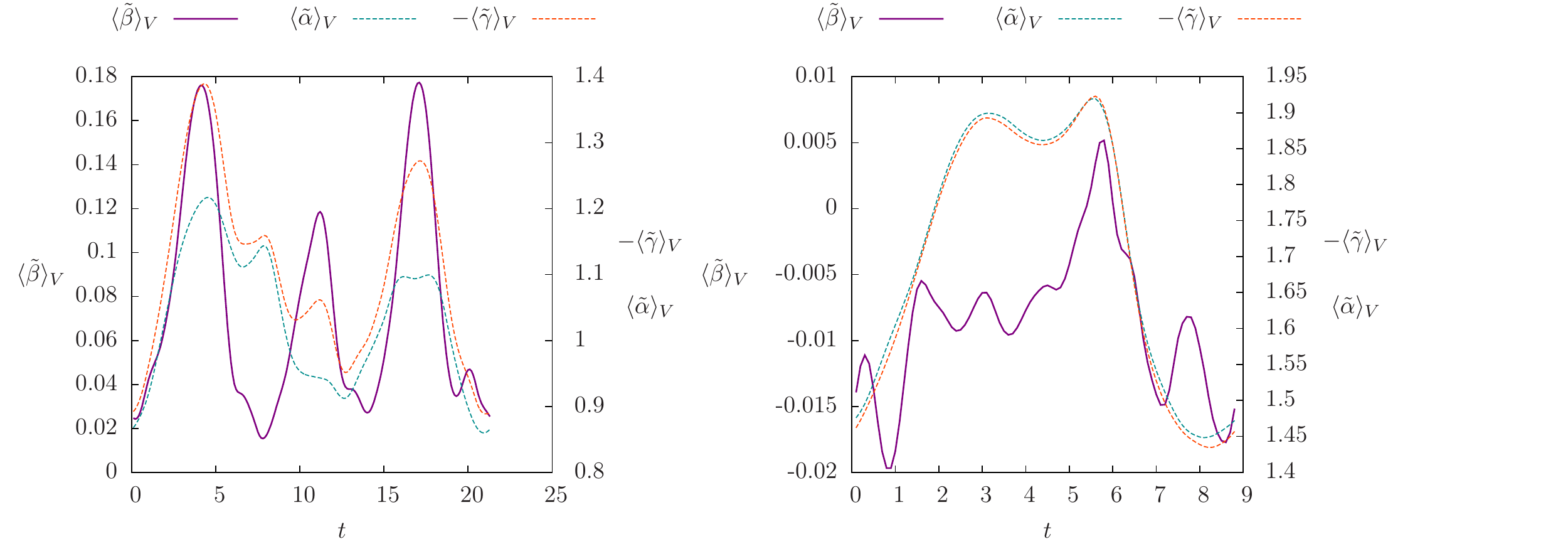}
\caption{\label{fig:strain}Time series of the volume averaged principal strains for UPO-o1 (left) and UPO-l1 (right). For the overturning case, peaks of $\mathscr E$ are coincident with large positive intermediate strain $\langle \tilde \beta\rangle_V,$ nearer to the values of isotropic turbulence, where as the more layered case (right) is very anisotropic with small $\langle \tilde \beta\rangle_V$, approximately zero when $\mathscr E$ is largest.}
\end{figure}

{In order to characterise the straining/scouring motions exhibited by the `l' class, we examine the properites of the strain tensor
$$ S_{ij}=\frac{1}{2}\left( \frac{\partial u_i}{\partial x_j}+\frac{\partial u_j}{\partial x_i}\right).$$
This tensor has three real eigenvalues, which we denote by $\tilde{\alpha},\,\tilde{\beta},\,\tilde{\gamma},$ referred to as the principal strains, which sum to zero in an incompressible flow. In general $\tilde{\alpha}>0$ and is therefore the stretching strain, $\tilde{\gamma}<0$ and is the compressional strain and the intermediate strain $\tilde{\beta}$ will control the three dimensionality, i.e.  $\tilde{\beta}<0$ represent compression in two directions and fluid elements drawn into filaments, $\tilde{\beta}>0$ compression in one direction and fluid elements are drawn into sheets and $\tilde{\beta}=0$ a purely two-dimensional straining field. Figure \ref{fig:strain} shows time series of the volume-averaged strains for the two characteristic UPOs, UPO-o1 and UPO-l1. In the overturning case, UPO-o1 shows that $\langle \tilde \beta\rangle_V$ is well correlated to the mixing efficiency, being positive and maximum when $\mathscr E$ is largest. This is in good agreement with \cite{SMYTH:1999io} who examined these strains for a freely decaying Kelvin-Helmholtz unstable shear layer;  $\langle \tilde \beta\rangle_V$ is positive and has a distinct maximum during the second phase of the overturn where the flow becomes more isotropic (isotropic turbulence is observed to have the strains approximately in the ratio 3:1:-4 \citep{SMYTH:1999io,Ashurst:1987}) and approaches zero in the quiescent periods as the flow relaxes back to the parallel shear mean flow. For the layered case, UPO-l1 is distinctly more anisotropic; the overall absolute values for  $\langle \tilde \beta\rangle_V$ are an order of magnitude smaller and are actually negative for much of the cycle. At the point where the mixing efficiency is maximal ($t\approx5.5$) $\langle \tilde \beta\rangle_V$ also has a maximum, but here this amounts to crossing the axes and $\langle \tilde \alpha\rangle_V\approx-\langle \tilde \gamma\rangle_V.$ Our interpretation is then that the flow approaches two-dimensionality where the mixing is largest and increased vertical gradients of $\rho$ are formed by the change in sign of  $\langle \tilde \beta\rangle_V.$ This is in stark contrast to the overturning case where mixing has a more isotropic straining signature with large $\langle \tilde \beta\rangle_V.$ As a secondary observation, the higher frequency oscillations of $\langle \tilde \beta\rangle_V$ for UPO-l1 are of the order of $N=\sqrt{B}$ the buoyancy frequency, i.e. the buoyancy period here is $T_N=2\pi/\sqrt{50}\approx0.9.$ This suggests that internal waves are also playing a role in this dynamical cycle}. 

\section{Discussion and Conclusions}

In this paper we have successfully applied recurrent flow analysis to stratified flows for the first time. As expected, success is restricted to relatively weak turbulence at modest Reynolds numbers. Nevertheless, the approach has still provided detailed insight into the sustaining mechanisms and mixing processes at work in these flows. In particular we have established that mixing can occur from two distinct mechanisms; overturning and scouring, consistently with the classification
presented by \cite{woodsjfm2010}. Scouring is observed when stratification is strong and overturns when the stratification is weak enough to allow small gradient Richardson numbers. By examining the cycles in space and time we have demonstrated that the spatial distribution of local gradient $Ri_G$ is markedly skewed before high overturning mixing,
but remains bounded from below by  $1/4$ when scouring mixing dominates.

From this analysis it is clear that characterising mixing by appropriate measures of Richardson number is important from both a statistical and dynamical point of view, although we have also shown that weak mixing can be controlled by processes uncorrelated to $Ri_G$. Crucially,  weaker average mixing may not be simply controlled by spatiotemporally intermittent shear instability, but by other scouring, straining 
processes, which may require
other predictive diagnostics. 
Furthermore, since the bulk measures of Richardson number and buoyancy Reynolds number are
functionally related, the results of our motivational DNS for horizontally forced, sustained turbulence
suggest
that an $(\overline{Ri}_B, \overline{Re}_B)$ parameterisation of mixing 
should be treated with caution, particularly since it proved difficult to access a strongly
stratified, yet strongly turbulent regime, analogously to 
the situation arising in stratified plane Couette flow \citep{Zhou:2017dd}.

One open question
 is the interplay between scouring and overturning. Clearly overturning shear instability will overwhelm any background straining or scouring motions, however it remains a challenge of rationalising spatiotemporal chaos to predict the switching between these processes. For instance simulation B3 shows that intermittent bursts can raise the local (in time) mixing efficiency, despite the majority of the mixing in this case being controlled by the straining and scouring layered motions characteristic of class `l'. Examination of a segment of the trajectory reaching large $\mathscr{E}$ in this case suggests straining and shear instability, saturating at various finite amplitudes, combine to populate the distribution of figure \ref{fig:mix_ReB}(b). Furthermore, as already widely discussed,
research into exact coherent structures embedded in turbulent flows faces a serious challenge in the extension of 
methods to handle higher Reynolds numbers and spatially localised dynamics. It remains unclear whether such UPOs
can be identified in `energetic' stratified turbulence with $\overline{Re}_B \gtrsim O(100)$, yet it is of particular
interest to understand whether the observed  reduction of mixing efficiency at such high $\overline{Re}_B$ is 
associated with a qualitative change in mixing properties. 

\vspace{2cm}
\noindent
{\em Acknowledgements}. 
We extend our thanks, for many helpful and enlightening discussions, to Paul Linden, John Taylor, Stuart Dalziel and the rest of the `MUST' team in Cambridge and Bristol. We also thank the three anonymous referees whose constructive comments have significantly improved the clarity of the manuscript. The source code used in this work is provided at \url{https://bitbucket.org/dan_lucas/psgpu} and the associated data including initialisation files and converged states can be found at \url{www.repository.cam.ac.uk} This work is supported by EPSRC Programme Grant EP/K034529/1 entitled `Mathematical Underpinnings of Stratified Turbulence'. The majority of the research presented here was conducted when DL was a postdoctoral researcher in DAMTP as part of the MUST programme grant.

\bibliography{papers_SKF2}

\begin{thebibliography}{24}
\expandafter\ifx\csname natexlab\endcsname\relax\def\natexlab#1{#1}\fi
\def\au#1{#1} \def\ed#1{#1} \def\yr#1{#1}\def\at#1{#1}\def\jt#1{\textit{#1}}
  \def\bt#1{#1}\def\bvol#1{\textbf{#1}} \def\vol#1{#1} \def\pg#1{#1}
  \def\publ#1{#1}\def\arxiv#1{#1}\def\org#1{#1}\def\st#1{\textit{#1}}

\bibitem[Ashurst {\em et~al.\/}(1987)Ashurst, Kerstein, Kerr \&
  Gibson]{Ashurst:1987}
{\sc \au{Ashurst, Wm.~T.}, \au{Kerstein, A.~R.}, \au{Kerr, R.~M.} \&
  \au{Gibson, C.~H.}} \yr{1987}  \at{Alignment of vorticity and scalar gradient
  with strain rate in simulated navier–stokes turbulence}.  \jt{Phys. Fluids}
   \bvol{30}~(8),  \pg{2343--2353}.

\bibitem[Chandler \& Kerswell(2013)]{Chandler:2013fi}
{\sc \au{Chandler, G.~J.} \& \au{Kerswell, R.~R.}} \yr{2013}  \at{{Invariant
  recurrent solutions embedded in a turbulent two-dimensional Kolmogorov
  flow}}.  \jt{J. Fluid Mech.}  \bvol{722},  \pg{554--595}.

\bibitem[Cvitanovi{\'c} \& Gibson(2010)]{2010PhST..142a4007C}
{\sc \au{Cvitanovi{\'c}, P.} \& \au{Gibson, J.~F.}} \yr{2010}  \at{{Geometry of
  the turbulence in wall-bounded shear flows: periodic orbits}}.  \jt{Physica
  Scripta}  \bvol{142},  \pg{4007}.

\bibitem[Garaud {\em et~al.\/}(2015)Garaud, Gallet \& Bischoff]{Garaud:2015ce}
{\sc \au{Garaud, Pascale}, \au{Gallet, Basile} \& \au{Bischoff, Tobias}}
  \yr{2015}  \at{{The stability of stratified spatially periodic shear flows at
  low P{\'e}clet number}}.  \jt{Physics of Fluids}  \bvol{27}~(8),
  \pg{084104--22}.

\bibitem[Ivey {\em et~al.\/}(2008)Ivey, Winters \& Koseff]{Ivey:2008hm}
{\sc \au{Ivey, G.~N.}, \au{Winters, K.~B.} \& \au{Koseff, J.~R.}} \yr{2008}
  \at{{Density stratification, turbulence, but how much mixing?}}  \bt{In {\em
  Annu. Rev. Fluid Mech.\/}}, ,  \vol{vol.~40},  \pg{pp. 169--184}.

\bibitem[Kawahara \& Kida(2001)]{Kawahara:2001ft}
{\sc \au{Kawahara, G.} \& \au{Kida, S.}} \yr{2001}  \at{{Periodic motion
  embedded in plane Couette turbulence: regeneration cycle and burst}}.  \jt{J.
  Fluid Mech.}  \bvol{449},  \pg{291}.

\bibitem[Kawahara {\em et~al.\/}(2012)Kawahara, Uhlmann \& {van
  Veen}]{Kawahara:2012iu}
{\sc \au{Kawahara, G.}, \au{Uhlmann, M.} \& \au{{van Veen}, L.}} \yr{2012}
  \at{{The significance of simple invariant solutions in turbulent flows}}.
  \jt{Annu. Rev. Fluid Mech.}  \bvol{44}~(1),  \pg{203--225}.

\bibitem[Linden(1979)]{linden1979}
{\sc \au{Linden, P.~F.}} \yr{1979}  \at{{Mixing in stratified fluids}}.
  \jt{Geophys. Astrophys. Fluid Dyn.}  \bvol{13}~(1),  \pg{3--23}.

\bibitem[Lucas {\em et~al.\/}(2017)Lucas, Caulfield \& Kerswell]{Lucas:2017wc}
{\sc \au{Lucas, D.}, \au{Caulfield, C.~P.} \& \au{Kerswell, R.~R.}} \yr{2017}
  \at{{Layer formation in horizontally forced stratified turbulence: connecting
  exact coherent structures to linear instabilities}} ,  \arxiv{arXiv:
  1701.05406v1}.

\bibitem[Lucas \& Kerswell(2015)]{Lucas:2015gt}
{\sc \au{Lucas, D.} \& \au{Kerswell, R.~R.}} \yr{2015}  \at{{Recurrent flow
  analysis in spatiotemporally chaotic 2-dimensional Kolmogorov flow}}.
  \jt{Phys. Fluids}  \bvol{27}~(4),  \pg{045106--27}.

\bibitem[Lucas \& Kerswell(2017)]{Lucas:2017fz}
{\sc \au{Lucas, D.} \& \au{Kerswell, R.~R.}} \yr{2017}  \at{{Sustaining
  processes from recurrent flows in body-forced turbulence}}.  \jt{Journal of
  Fluid Mechanics}  \bvol{817},  \pg{R3--11}.

\bibitem[Maffioli {\em et~al.\/}(2016)Maffioli, Brethouwer \&
  Lindborg]{Maffioli:2016bw}
{\sc \au{Maffioli, A.}, \au{Brethouwer, G.} \& \au{Lindborg, E.}} \yr{2016}
  \at{{Mixing efficiency in stratified turbulence}}.  \jt{J. Fluid Mech.}
  \bvol{794},  \pg{R3--12}.

\bibitem[Mashayek {\em et~al.\/}(2017)Mashayek, Salehipour, Bouffard,
  Caulfield, Ferrari, Nikurashin, Peltier \& Smyth]{mashayekgrl2017}
{\sc \au{Mashayek, A.}, \au{Salehipour, H.}, \au{Bouffard, D.}, \au{Caulfield,
  C.~P.}, \au{Ferrari, R.}, \au{Nikurashin, M.}, \au{Peltier, W.~R.} \&
  \au{Smyth, W.~D.}} \yr{2017}  \at{Efficiency of turbulent mixing in the
  abyssal ocean circulation}.  \jt{Geophys. Res. Lett.} 10.1002/2016GL072452.

\bibitem[Mater \& Venayagamoorthy(2014)]{matergrl2014}
{\sc \au{Mater, B.~D.} \& \au{Venayagamoorthy, S.~K.}} \yr{2014}  \at{The quest
  for an unambiguous parameterization of mixing efficiency in stably stratified
  geophysical flows}.  \jt{Geophys. Res. Lett.}  \bvol{41}~(13),
  \pg{4646--4653}.

\bibitem[Osborn(1980)]{Osborn:1980jj}
{\sc \au{Osborn, T.~R.}} \yr{1980}  \at{{Estimates of the Local Rate of
  Vertical Diffusion from Dissipation Measurements}}.  \jt{Journal of Physical
  Oceanography}  \bvol{10}~(1),  \pg{83--89}.

\bibitem[Peltier \& Caulfield(2003)]{Peltier:2003gt}
{\sc \au{Peltier, W.~R.} \& \au{Caulfield, C.~P.}} \yr{2003}  \at{{Mixing
  efficiency in stratified shear flows}}.  \jt{Annu. Rev. Fluid Mech.}
  \bvol{35},  \pg{135--167}.

\bibitem[Salehipour \& Peltier(2015)]{Salehipour:2015bt}
{\sc \au{Salehipour, H.} \& \au{Peltier, W.~R.}} \yr{2015}  \at{{Diapycnal
  diffusivity, turbulent Prandtl number and mixing efficiency in Boussinesq
  stratified~turbulence}}.  \jt{J. Fluid Mech.}  \bvol{775},  \pg{464--500}.

\bibitem[Salehipour {\em et~al.\/}(2016)Salehipour, Peltier, Whalen \&
  MacKinnon]{HSalehipour:2016jr}
{\sc \au{Salehipour, H.}, \au{Peltier, W.~R.}, \au{Whalen, C.~B.} \&
  \au{MacKinnon, J.~A.}} \yr{2016}  \at{{A new characterization of the
  turbulent diapycnal diffusivities of mass and momentum in the ocean}}.
  \jt{Geophys. Res. Lett.}  \bvol{43}.

\bibitem[Scotti \& White(2016)]{scottijpo2016}
{\sc \au{Scotti, A.} \& \au{White, B.}} \yr{2016}  \at{The mixing efficiency of
  stratified turbulent boundary layers}.  \jt{J. Phys. Oceanogr.}
  \bvol{46}~(10),  \pg{3181--3191}.

\bibitem[Shih {\em et~al.\/}(2005)Shih, Koseff, Ivey \& Ferziger]{SHIH:2005bx}
{\sc \au{Shih, L.~H.}, \au{Koseff, J.~R.}, \au{Ivey, G.~N.} \& \au{Ferziger,
  J.~H.}} \yr{2005}  \at{{Parameterization of turbulent fluxes and scales using
  homogeneous sheared stably stratified turbulence simulations}}.  \jt{J. Fluid
  Mech.}  \bvol{525},  \pg{193--214}.

\bibitem[Smyth(1999)]{SMYTH:1999io}
{\sc \au{Smyth, W.~D.}} \yr{1999}  \at{{Dissipation-range geometry and scalar
  spectra in sheared stratified turbulence}}.  \jt{J. Fluid Mech.}  \bvol{401},
   \pg{209--242}.

\bibitem[{van Veen} {\em et~al.\/}(2006){van Veen}, Kida \&
  Kawahara]{vanVeen:2006fm}
{\sc \au{{van Veen}, L.}, \au{Kida, S.} \& \au{Kawahara, G.}} \yr{2006}
  \at{{Periodic motion representing isotropic turbulence}}.  \jt{Fluid Dyn.
  Res.}  \bvol{38}~(1),  \pg{19--46}.

\bibitem[Woods {\em et~al.\/}(2010)Woods, Caulfield, Landel \&
  A.]{woodsjfm2010}
{\sc \au{Woods, A.~W.}, \au{Caulfield, C.~P.}, \au{Landel, J.~R.} \& \au{A.,
  Kuesters}} \yr{2010}  \at{Non-invasive turbulent mixing across a density
  interface in a turbulent taylor-couette flow}.  \jt{J. Fluid Mech.}
  \bvol{663},  \pg{347--357}.

\bibitem[Zhou {\em et~al.\/}(2017)Zhou, Taylor \& Caulfield]{Zhou:2017dd}
{\sc \au{Zhou, Q.}, \au{Taylor, J.~R.} \& \au{Caulfield, C.~P.}} \yr{2017}
  \at{{Self-similar mixing in stratified plane {C}ouette flow for varying
  Prandtl number}}.  \jt{J. Fluid Mech.}  \bvol{820},  \pg{86--120}.

\end{thebibliography}
\bibliographystyle{jfm}

\end{document}